\documentclass[osajnl,twocolumn,showpacs]{revtex4}

\usepackage{graphicx}
\graphicspath{{pict/}{}}

\newcommand\pictc[5]{\begin{figure}
            \centerline{\vspace{0mm}
\includegraphics[width=\columnwidth,height=0.7\textheight,keepaspectratio]{#3}}
            \protect\caption{\protect\label{fig:#4} #5}
                    \end{figure}            }

\newcommand\pict[4][1]{\pictc{#1}{!tb}{#2}{#3}{#4}}
\newcommand\rpict[1]{\ref{fig:#1}}

\newcounter{Fig}

\begin{document}
\begin{sloppy}

\title{Slow-light switching in nonlinear Bragg-grating coupler}

\author{Sangwoo Ha}
\author{Andrey A. Sukhorukov}
\author{Yuri S. Kivshar}

\affiliation{Nonlinear Physics Centre and Centre for Ultra-high bandwidth Devices for Optical Systems (CUDOS),\\
 Research School of Physical Sciences and Engineering, Australian National University, Canberra, ACT 0200, Australia}

\begin{abstract}
We study propagation and switching of slow-light pulses in nonlinear
couplers with phase-shifted Bragg gratings. We demonstrate that
power-controlled nonlinear self-action of light can be used to
compensate dispersion-induced broadening of pulses through the formation of
gap solitons, to control pulse switching in the coupler, and to tune the
propagation velocity.
\end{abstract}


\maketitle

It is known that the speed of light can be dramatically reduced in
photonic-crystal waveguides with a periodic modulation of the
optical refractive index~\cite{Vlasov:2005-65:NAT,
Gersen:2005-73903:PRL, Jacobsen:2005-7861:OE}. In the regime of slow
light the photon-matter interactions are dramatically
enhanced~\cite{Soljacic:2002-2052:JOSB} allowing for all-optical
control and manipulation. In particular, nonlinear self-action can
be used to dynamically tune the velocity of pulses propagating in
nonlinear Bragg gratings~\cite{Mok:2006-775:NAPH}. At the same time,
nonlinearity can support pulse self-trapping in the form of a gap
soliton which profile remains undistorted during propagation~\cite{Mok:2006-775:NAPH}, overcoming the limitations of linear devices due to dispersion-induced pulse broadening~\cite{Engelen:2006-1658:OE}.

Nonlinear effects can also enable ultra-fast all-optical pulse
switching. Pulse routing between output ports controlled by optical
power was demonstrated in directional
couplers~\cite{Jensen:1982-1568:ITMT, Maier:1982-2296:KE,
Friberg:1987-1135:APL}. Additional flexibility in mode conversion
with applications to add-drop filters~\cite{Orlov:1997-688:OL,
Perrone:2001-1943:JLT, Tomljenovic-Hanic:2005-1615:JOSA} is realized
by combining directional coupler geometry and Bragg gratings in
optical fibers~\cite{Aslund:2003-6578:AOP} or planar photonic
structures~\cite{Castro:2006-1236:AOP} which operation can be tuned
all-optically~\cite{Imai:2005-293:IPAPB}, and gap solitons can also
exist in the nonlinear regime~\cite{Mak:1998-1685:JOSB,
Mak:2004-66610:PRE, Gubeskys:2004-283:EPD}.

In this Letter, we suggest novel possibilities for dynamic
manipulation of slow-light pulses which frequency is tuned in the vicinity of Bragg resonance. The pulse dynamics can be modeled by a set of  coupled-mode nonlinear equations~\cite{Agrawal:1989:NonlinearFiber} for the
normalized slowly varying envelopes of the forward ($u_n$) and backward ($w_n$) propagating fields in each of waveguides $n=1,2$,
$      -i {\partial u_n}/{\partial t} =
     i {\partial u_n}/{\partial z}
      + C u_{3-n}
      + \rho_n w_n
      + \gamma (|u_n|^2 + 2 |w_n|^2) u_n$,
$      -i {\partial w_n}/{\partial t}
     = -i {\partial w_n}/{\partial z}
      + C w_{3-n}
      + \rho_n^\ast u_n
      + \gamma (|w_n|^2 + 2 |u_n|^2) w_n$,
where $t$ and $z$ are the dimensionless time and propagation
distance normalized to $t_s$ and $z_s$, respectively, $C$ is the
coupling coefficient for the modes of the neighboring waveguides,
$\rho_n$ characterizes the amplitude and phase of the Bragg
gratings, $\gamma$ is the nonlinear coefficient, and the group
velocity far from the Bragg resonance is normalized to unity. The
scaling coefficients are $t_s = \lambda_0^2 |\rho_1| / (\pi c
\Delta \lambda_0)$ and $z_s = t_s c / n_0$, where $c$ is the speed
of light in vacuum, $\lambda_0$ is the wavelength in vacuum, $\Delta
\lambda_0$ is the width of Bragg resonance for an individual
waveguide, $n_0$ is the effective refractive index in the absence of
a grating. To be specific, in numerical examples we set
$\gamma=10^{-2}$, $\lambda_0 = 1550.63 nm$, $\Delta \lambda_0 = 0.1$nm, $t_s \simeq 12.8 ps$, $z_s \simeq 1.8 mm$ corresponding to characteristic parameters of fiber Bragg
gratings~\cite{Aslund:2003-6578:AOP, Mok:2006-775:NAPH}.

\pict{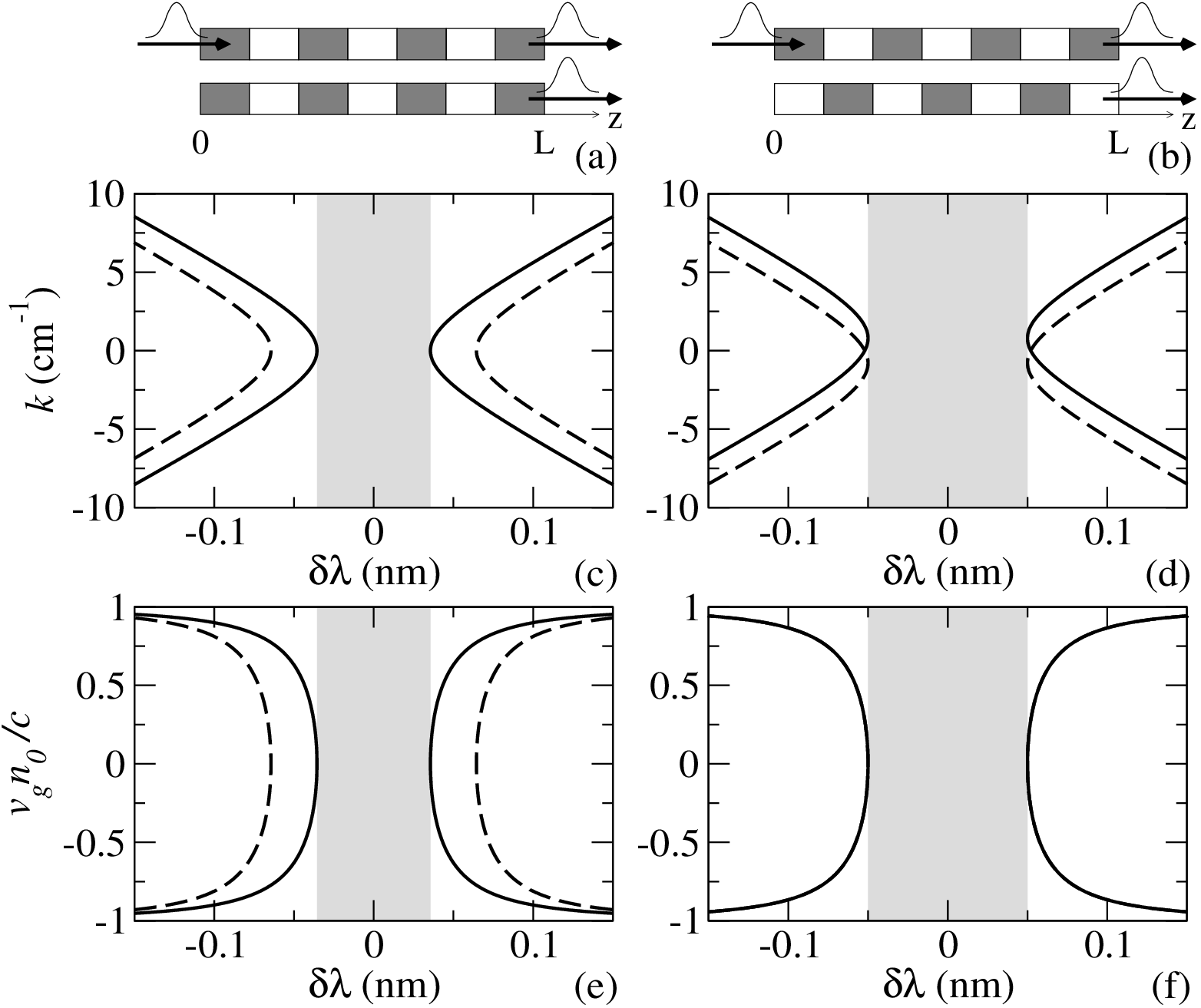}{couplerModes}{ (a,b)~Schematic of directional
couplers with (a)~in-phase ($\rho_1 = \rho_2=0.5$) or
(b)~out-of-phase ($\rho_1 = -\rho_2 = 0.5$) Bragg gratings.
(c,d)~Characteristic dispersion and (e,f)~normalized group velocity
dependence on wavelength detuning for the case of in-phase~(c,e) and
out-of-phase~(d,f) gratings. For all the plots $C \simeq 0.144$. }

We consider the case of identical waveguides and analyze the
effect of {\em a phase shift} ($\varphi$) between the otherwise
equivalent waveguide gratings with $\rho_1 = \rho$ and $\rho_2 =
\rho \exp(i \varphi)$ (with no loss of generality, we take $\rho$ to
be real and positive), see schematic illustrations in
Figs.~\rpict{couplerModes}(a,b). It was shown that the grating shift
can strongly modify the reflectivity of modes with different
symmetries~\cite{Perrone:2001-1943:JLT, Aslund:2003-6578:AOP,
Castro:2006-1236:AOP}, and we investigate how this structural
parameter affects the properties of slow-light pulses.

In the linear regime, wave propagation is fully defined through the Floquet-Bloch eigenmode solutions of the form,
      $u_n = U_n \exp\left( i \beta z - i \omega t \right)$,
      $w_n = W_n \exp\left( i \beta z - i \omega t \right)$.
After substituting these expressions into the linearized coupler
equations (with $\gamma=0$), we obtain the dispersion relation
$\omega^2(\beta) = \beta^2 + C^2 + |\rho|^2 \pm 2 C [ \beta^2 +
|\rho|^2 \cos^2(\varphi/2)]^{1/2}$.

Slow-light propagation can be observed due to the reduction of
the normalized group velocity ($v_g = d \omega / d \beta$) when the
pulse frequency is tuned close to the bandgap edge, where the
propagating waves with real $\beta$ are absent. We find that
different regimes of slow light can be realized depending on the
structural parameters. (i)~If $|\rho \cos(\varphi/2) / C| > 1$, the
bandgap appears for $\omega^2 < \omega_g^2 = C^2 + |\rho|^2 - 2 C
|\rho \cos(\varphi/2)|$, and only a single forward propagating mode
(with $v_g >0$) exists for the frequencies near the gap edges, see
examples in Figs.~\rpict{couplerModes}(c,e). (ii)~If $|\rho
\cos(\varphi/2) / C| < 1$, the bandgap appears for $|\omega| <
\omega_g = |\rho \sin(\varphi/2)|$, and two types of the forward
propagating modes (with $v_g >0$) exist simultaneously (in the
regions with $\beta>0$ and $\beta<0$) for the frequencies
arbitrarily close to the gap edges, see examples in
Figs.~\rpict{couplerModes}(d,f).

\pict{fig02}{couplerTransmit}{ Linear transmission of
incident wave coupled to the first waveguide of a semi-infinite ($z
\ge 0$) coupler with (a,c,e)~in-phase and (b,d,f)~out-of-phase
Bragg gratings: (a-d)~Intensity distribution (averaged over grating
period) shown in the first (solid line) and second (dashed line)
waveguides for (a,b)~large frequency detuning from the resonance and
(c,d)~frequency tuned close to the band edge with slow group
velocity $v_g = 0.1$. (e,f)~Intensities at $z = 2$cm vs. wavelength
detuning. Parameters correspond to Fig.~\rpict{couplerModes}, and
the intensities are normalized to the input intensity. }

We now analyze linear propagation of pulses in a
semi-infinite Bragg grating coupler. When the optical frequency is
detuned from the bandgap, light periodically tunnels between the
waveguides with the characteristic period $L_c \simeq \pi / (2 C)$
defined for a conventional coupler without the Bragg grating, see
examples in Figs.~\rpict{couplerTransmit}(a,b). The periodic
tunneling appears due to the beating of even and odd modes, which
correspond to different branches of the dispersion curves. When the
pulse frequency is tuned closer to the gap edge and (i)~only one
slow mode is supported, then periodic beating disappears and light
is equally distributed between the waveguides irrespective of the
input excitation, see Figs.~\rpict{couplerTransmit}(c,e). The
periodic coupling can only be sustained in the slow-light regime
when (ii)~two modes co-exist at the gap edge, see
Figs.~\rpict{couplerTransmit}(d,f). Therefore, the configuration
with {\em out-of-phase shifted Bragg gratings is the most
preferential} for switching of slow-light pulses, since for
$\varphi=\pi$ the dispersion of the type (ii) is always realized for any
values of the grating strength and the waveguide coupling, and
simultaneously the bandgap attains the maximum bandwidth.

\pict{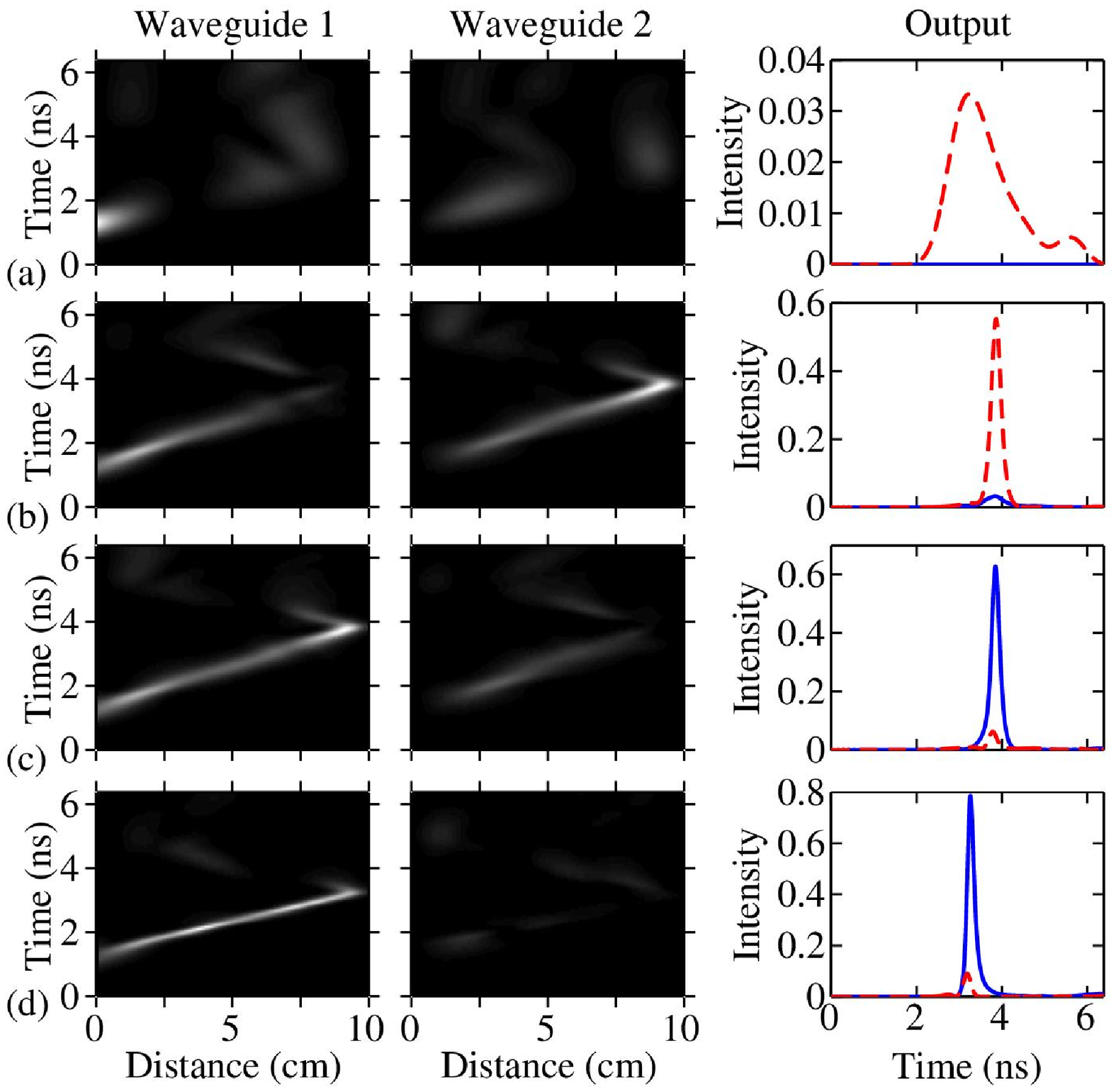}{couplerDynamics}{ (a-d)~Pulse dynamics inside
the nonlinear coupler for different values of the normalized peak
input intensities $I_0 = 10^{-4}, 3.33, 3.37, 4$. Shown are the
density plots of intensity in the first (left column) and second
(middle column) waveguides. Output intensity profiles normalized to
$I_0$ at the first (solid line) and second (dashed line) waveguides
are shown in the right column. Input Gaussian pulse has full width at half-maximum of intensity of $577 ps$, and its central wavelength is tuned to the gap edge at $\lambda_0 - \Delta \lambda_0 / 2$.}

\pict{fig04}{couplerOutput}{ 
Dependence of output pulse characteristics on the input peak intensity: (a)~output power normalized to the input power; (b)~pulse full-width at half-maximum of intensity, dotted line marks the input pulse width; (c)~pulse delay relative to propagation without the Bragg grating normalized to the input pulse width. In all the plots, solid and dashed lines correspond to the outputs at the first and second waveguides,
respectively. }

At higher optical powers, nonlinear effects become important, and we
perform numerical simulations of the coupler equations to model
pulse propagation. Examples of the pulse dynamics
and output pulse characteristics are presented in
Figs.~\rpict{couplerDynamics} and~\rpict{couplerOutput}, where we
consider the structure size equal to three coupling lengths, $L =
3 L_c$. In the linear regime, the pulse tunnels three times between
the waveguides and switches accordingly to the other waveguide at
the output, see Fig.~\rpict{couplerDynamics}(a). However, at the
same time the pulse significantly broadens due to the effect of
the group-velocity dispersion (GVD). As the input pulse energy is
increased, nonlinearity may support dispersionless slow-light pulses
in the form of gap solitons, studied previously in
single~\cite{Mok:2006-775:NAPH} and coupled waveguides with in-phase
gratings~\cite{Mak:1998-1685:JOSB, Mak:2004-66610:PRE,
Gubeskys:2004-283:EPD}. Most remarkably, we find that the presence
of two types of slow-light modes in the structure with out-of-phase
gratings gives rise to a {new type of gap solitons} which
periodically tunnel between the waveguides while preserving a
constant width, see Figs.~\rpict{couplerDynamics}(b-d) and
Fig.~\rpict{couplerOutput}(b). In agreement with the properties of
conventional nonlinear couplers~\cite{Jensen:1982-1568:ITMT,
Maier:1982-2296:KE, Friberg:1987-1135:APL}, the coupling length is
gradually extended as the optical power is increased, resulting in
the pulse switching between the output waveguides, see
Fig.~\rpict{couplerOutput}(a). As the input power is further
increased, we observe a {\em sharp switching} when the output is highly sensitive to small changes of the input intensity (less than 1\%), cf. Figs.~\rpict{couplerDynamics}(b) and~(c). At the same time, the pulse delay is also varied with optical power, as shown in Fig.~\rpict{couplerOutput}(c). The power
tunability of the pulse delay and switching dynamics can be adjusted by
selecting parameters such as waveguide coupling, and choosing the
frequency detuning from the gap edge.

In conclusion, we have demonstrated that flexible manipulation of
slow-light pulses can be realized in a nonlinear couplers with
phase-shifted Bragg gratings, implemented as all-fiber~\cite{Aslund:2003-6578:AOP} or planar waveguide devices created in highly nonlinear materials~\cite{Millar:1999-685:OL, Shokooh-Saremi:2006-1323:JOSB}. We predict the possibility to simultaneously suppress pulse spreading due to dispersion, all-optically tune the pulse velocity and transit delays, and switch pulses between the
output ports. We anticipate that similar effects may be achieved in
other types of photonic structures including photonic-crystal
waveguides~\cite{Mori:2005-9398:OE} and
fibers~\cite{Ibanescu:2004-63903:PRL} engineered to
support several co-propagating slow-light modes. Our
results also suggest new opportunities for control of slow-light
bullets in periodic waveguide
arrays~\cite{Sukhorukov:2006-press:PRL}.

We thank M. de Sterke  and B.~Eggleton for useful discussions. This work has been supported by the Australian Research Council.

\end{sloppy}
\end{document}